\documentclass[twocolumn]{aastex63}

\usepackage{textcomp, gensymb, amsmath}

\shorttitle{Helium around WASP-69b and WASP-52b}
\shortauthors{Vissapragada et al.}

\begin{document}

\title{Constraints on Metastable Helium in the Atmospheres of WASP-69b and WASP-52b with Ultra-Narrowband Photometry}

\correspondingauthor{Shreyas Vissapragada}
\email{svissapr@caltech.edu}

\author[0000-0003-2527-1475]{Shreyas Vissapragada}
\affiliation{Division of Geological and Planetary Sciences, California Institute of Technology, 1200 East California Blvd, Pasadena, CA 91125, USA}

\author{Heather A. Knutson}
\affil{Division of Geological and Planetary Sciences, California Institute of Technology, 1200 East California Blvd, Pasadena, CA 91125, USA}

\author[0000-0001-5213-6207]{Nemanja Jovanovic}
\affil{Department of Astronomy, California Institute of Technology, 1200 East California Blvd, Pasadena, CA 91125, USA}

\author[0000-0001-5737-1687]{Caleb K. Harada}
\affil{Department of Astronomy, University of Maryland, 4296 Stadium Drive, College Park, MD 20742, USA}
\affil{Center for Astrophysics $|$ Harvard \& Smithsonian, 60 Garden Street, MS-16, Cambridge, MA 02138, USA}

\author[0000-0002-9584-6476]{Antonija Oklop{\v{c}}i{\'c}}
\affil{Center for Astrophysics $|$ Harvard \& Smithsonian, 60 Garden Street, MS-16, Cambridge, MA 02138, USA}
\affil{NHFP Sagan Fellow}

\author{James Eriksen}
\affil{Palomar Observatory, California Institute of Technology, 35899 Canfield Rd, Palomar Mountain, CA 92060, USA}

\author[0000-0002-8895-4735]{Dimitri Mawet}
\affil{Department of Astronomy, California Institute of Technology, 1200 East California Blvd, Pasadena, CA 91125, USA}

\affil{Jet Propulsion Laboratory, California Institute of Technology, 4800 Oak Grove Dr, Pasadena, CA 91109, USA}

\author[0000-0001-6205-9233]{Maxwell A. Millar-Blanchaer}
\affil{Department of Astronomy, California Institute of Technology, 1200 East California Blvd, Pasadena, CA 91125, USA}

\author[0000-0002-1481-4676]{Samaporn Tinyanont}
\affil{Department of Astronomy, California Institute of Technology, 1200 East California Blvd, Pasadena, CA 91125, USA}

\author[0000-0002-1871-6264]{Gautam Vasisht}
\affil{Jet Propulsion Laboratory, California Institute of Technology, 4800 Oak Grove Dr, Pasadena, CA 91109, USA}

\begin{abstract}
Infrared observations of metastable 2$^3$S helium absorption with ground- and space-based spectroscopy are rapidly maturing, as this species is a unique probe of exoplanet atmospheres. Specifically, the transit depth in the triplet feature (with vacuum wavelengths near 1083.3 nm) can be used to constrain the temperature and mass loss rate of an exoplanet's upper atmosphere. Here, we present a new photometric technique to measure metastable 2$^3$S helium absorption using an ultra-narrowband filter (full-width at half-maximum of 0.635 nm) coupled to a beam-shaping diffuser installed in the Wide-field Infrared Camera (WIRC) on the 200-inch Hale Telescope at Palomar Observatory. We use telluric OH lines and a helium arc lamp to characterize refractive effects through the filter and to confirm our understanding of the filter transmission profile. We benchmark our new technique by observing a transit of WASP-69b and detect an excess absorption of $0.498\pm0.045$\% (11.1$\sigma$), consistent with previous measurements after considering our bandpass. Then, we use this method to study the inflated gas giant WASP-52b and place a 95th-percentile upper limit on excess absorption in our helium bandpass of 0.47\%. Using an atmospheric escape model, we constrain the mass loss rate for WASP-69b to be $5.25^{+0.65}_{-0.46}\times10^{-4}~M_\mathrm{J}/\mathrm{Gyr}$ ($3.32^{+0.67}_{-0.56}\times10^{-3}~M_\mathrm{J}/\mathrm{Gyr}$) at 7,000~K (12,000~K). Additionally, we set an upper limit on the mass loss rate of WASP-52b at these temperatures of $2.1\times10^{-4}~M_\mathrm{J}/\mathrm{Gyr}$ ($2.1\times10^{-3}~M_\mathrm{J}/\mathrm{Gyr}$). These results show that ultra-narrowband photometry can reliably quantify absorption in the metastable helium feature. 
\end{abstract}


\section{Introduction} \label{sec:intro}
Many of the currently known exoplanets are on short-period orbits and thus experience severe insolation. Such extreme environments can radically alter planetary evolution, potentially driving atmospheric mass loss via thermal escape \citep[e.g.][]{Tian15, Owen19}. Mass loss can in turn leave substantial imprints on observed planetary statistics, such as the dearth of planets between 1.5 and 2 Earth radii (the ``radius gap" or ``evaporation valley") and the so-called ``Neptune desert'' in the radius-period plane \citep{Lopez13, Owen13, Owen17, Fulton17, VanEylen18, Fulton18, Cloutier19, Hardegree-Ullman20}. Over the past two decades, most measurements of mass loss rates for close-in planets have been conducted at ultraviolet wavelengths, with Lyman-$\alpha$ detections for HD 209458b \citep{VidalMadjar03}, HD 189733b \citep{Lecavelier10, Lecavelier12}, GJ 436b \citep{Kulow14, Ehrenreich15, Lavie17}, and GJ 3470b \citep{Bourrier18}; tentative/marginal signals for TRAPPIST-1b and c \citep{Bourrier17a}, Kepler-444e and f \citep{Bourrier17b}, and K2-18b \citep{dosSantos20}; and non-detections for 55 Cnc e \citep{Ehrenreich12}, HD 97658b \citep{Bourrier17c}, GJ 1132 b \citep{Waalkes19}, and $\pi$ Men c \citep{GarciaMunoz20}. While in theory the large cross-section of this line should result in strong absorption during exoplanet transits, in practice geocoronal emission and interstellar absorption effectively mask out the line core for most stars, requiring these studies to study the absorption in the line's extended wings.

The neutral helium triplet (with vacuum wavelengths near 1083.3 nm) offers a way to circumvent the limitations of Lyman-$\alpha$ observations \citep{Seager00, Oklopcic18} by shifting to infrared wavelengths where both the Earth's atmosphere and the interstellar medium \citep[e.g.][]{Indriolo09} are effectively transmissive. \citet{Spake18} were the first to successfully observe an enhanced transit depth in He I for WASP-107b with Wide-Field Camera~3 (WFC3) on the \textit{Hubble Space Telescope} (\textit{HST}). Soon after, ground-based observations with the CARMENES high-resolution ($R\sim80,000$) spectrograph on the 3.5 m telescope at Calar Alto Observatory have confirmed the absorption signal and measured the He I line shape for HAT-P-11b \citep{Allart18} and WASP-107b \citep{Allart19}, and have additionally revealed excess helium absorption signals for HD 189733b \citep{Salz18}, HD 209458b \citep{AlonsoFloriano19}, and WASP-69b \citep{Nortmann18}. \textit{HST} WFC3 observations were also used to identify He I absorption for HAT-P-11b \citep{Mansfield18}, and recently Keck II/NIRSPEC and the Habitable-zone Planet Finder have observed helium in the atmospheres of WASP-107b \citep{Kirk20} and GJ 3470b \citep{Ninan19}, respectively. We note also the reported non-detections of helium in the atmospheres of KELT-9b, GJ 436b \citep[both][]{Nortmann18}, WASP-12b \citep{Kreidberg18}, GJ 1214b \citep{Crossfield19}, and K2-100b \citep{Gaidos20}. Due to its observational accessibility for ground- and space-based facilities, the helium triplet has been firmly established as a window into the upper atmospheres of exoplanets.

Here, we introduce ultra-narrowband helium photometry, a ground-based technique complementary to high-resolution spectroscopy that is specifically crafted to measure the helium absorption depth using an ultra-narrow bandpass filter. In this work, we benchmark our new technique on the Wide-field Infrared Camera (WIRC), at the prime focus of the Hale 200'' telescope at Palomar Observatory. We first measure the He I light curve of WASP-69b, a 1000~K, Saturn-mass, and Jupiter-size planet orbiting a K5 host star with $J = 8$ \citep{Anderson14}. We compare our results to those of \citet{Nortmann18}, and show that our results agree well with theirs. We then present the first He I light curve of the slightly warmer (1300~K), larger (1.27~$R_\mathrm{J}$), and heavier (0.46~$M_\mathrm{J}$) planet WASP-52b, which orbits a K2 host star with $J = 10.5$ \citep{Hebrard13}. In Section~\ref{sec:exp}, we detail the experimental design of our ultra-narrowband helium photometer. We discuss our observations and data reduction techniques in Section~\ref{sec:obs}. We present our results in Section~\ref{sec:res}, and conclude with a look towards future applications of ultra-narrowband photometry in Section~\ref{sec:conc}.

\section{Experimental Design} \label{sec:exp}
Our experiment is analogous to broad-band transit photometry performed previously \citep{Vissapragada20} with the Wide-field InfraRed Camera \citep[WIRC;][]{Wilson03} on the Hale 200" telescope at Palomar Observatory. The sole difference is that we use an ultra-narrowband filter (manufactured by Alluxa) that is centered on the helium feature. We used a combination of identifiable telluric OH emission lines as well as a helium lamp (naturally producing the feature in emission) to calibrate out refractive effects and ensure our knowledge of the filter transmission profile is accurate.

\subsection{Filter Properties}
Specifically, our filter has a center wavelength of 1083.3~nm in vacuum, at 77~K, and at an angle of incidence (AOI) of {7\degree}; a full width at half maximum (FWHM) of 0.635~nm; and a maximum transmission of 95.6\% (averaged across five positions on the filter). To cover the full spectral range to which our 2.5~$\mu$m cutoff Hawaii-II detector is sensitive, the filter also has OD4 absolute out-of-band blocking (i.e. a transmission less than 0.01\% everywhere outside the passband) from 500 to 3000~nm. We additionally utilize an Engineered Diffuser (located in a separate filter wheel from the helium filter) that molds the stellar point-spread functions (PSFs) into a top-hat shape with a FWHM of 3\arcsec. The diffuser increases observing efficiency and limits systematics related to PSF variations. When combined with our guiding software, which can keep pointing stable to within 2-3~pixels (equivalent to 0\farcs5-0\farcs75) over an entire night, this setup allows for powerful control of time-correlated systematics \citep{Stefansson17}. With this setup in place, we have recently demonstrated a precision of 0.16\% per 10 minute bin for $J = 14$ magnitude stars \citep{Vissapragada20}.

\begin{figure*}[ht!]
    \centering
    \includegraphics[width=\textwidth]{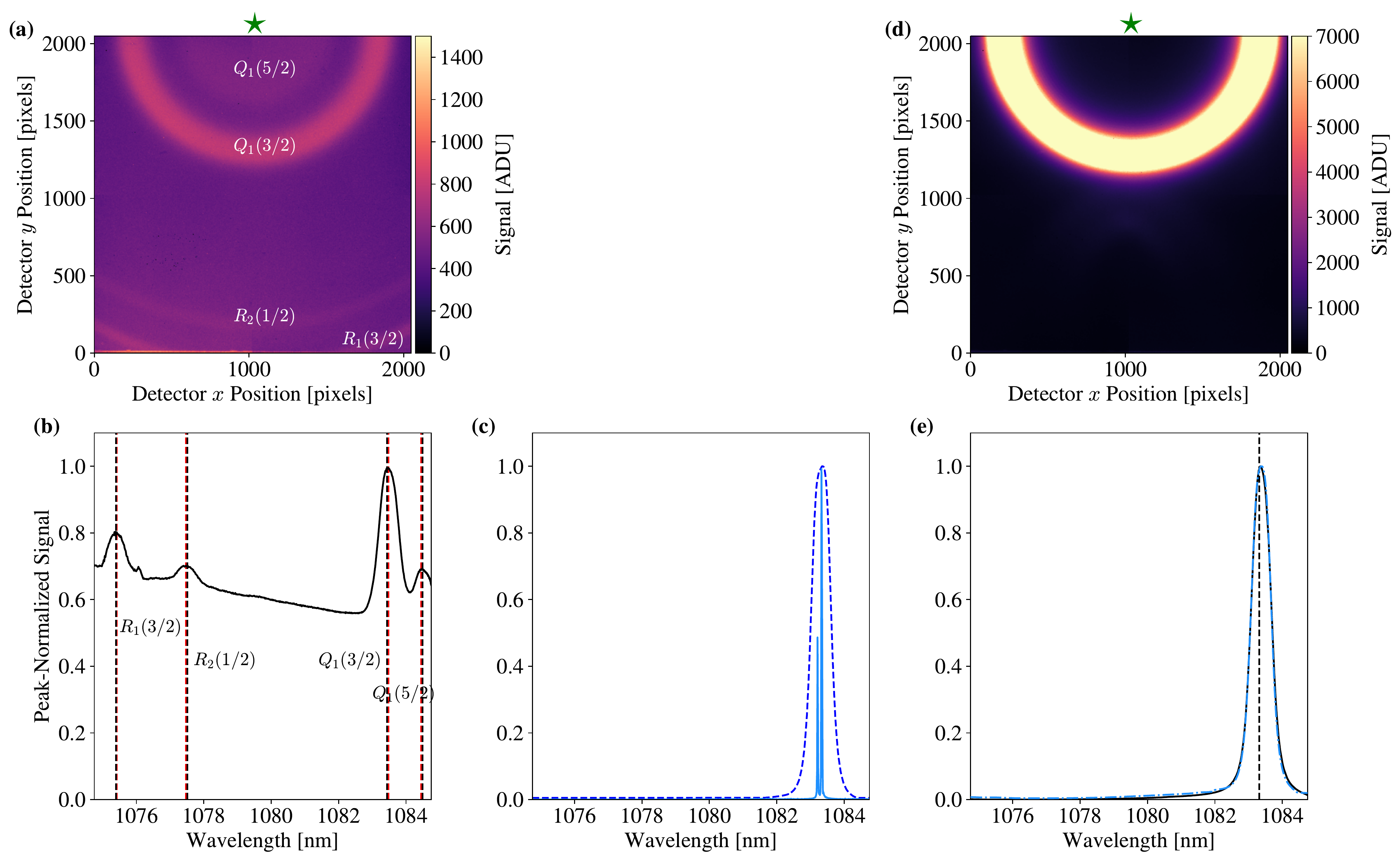}
    \caption{Experiment calibration. (a) 2048 by 2048 image of the sky background observed through WIRC and the helium filter. Telluric OH emission lines appear as arcs, and each strong line from the $\nu = 5-2$ band of ground-state OH is labeled. The green star indicates the zero point of the filter at $(x_0, y_0) = (1037, 2120)$. (b) The reconstructed spectrum of the sky from (a). Known positions of the telluric OH features are labeled and marked with black dashed lines, and line positions from the best-fit wavelength solution are marked with red dashed lines, which are effectively superimposed on the black dashed lines with small offsets. (c) Laboratory measurements of the helium lamp spectrum (light blue) and the helium filter transmission profile (dashed blue). (d) 2048 by 2048 image of the helium lamp observed through WIRC and the helium filter. The metastable helium triplet appears as a single bright arc due to convolution with the filter transmission profile. The green star again indicates the zero point of the filter. (e) The reconstructed spectrum of the helium arc lamp from (d), shown with a solid black line, compared to the laboratory spectrum of the helium arc lamp convolved with the filter transmission profile (dashed light blue) and the known wavelength of the feature (dashed black).}
    \label{fig1}
\end{figure*}

 Consideration of refractive effects is critical for such a narrowband filter, especially with a wide-field camera \citep[e.g.][]{Ghinassi02, Tinyanont19}. Critically, the filter wheels in WIRC are fixed at a {7\degree} tilt to minimize ghosting \citep{Wilson03}, and the filters cannot be angle-tuned. Because most rays forming the image encounter the filter at non-normal incidence due to the filter tilt (as well as the diversity of angles for each field point), they experience a different passband. As a result, different positions on the detector correspond to different filter transmission profiles. While this effect is noticeable even for broadband filters \citep{Ghinassi02, Tinyanont19}, the amplitude of the shift in wavelength space is small compared to the width of the bandpass, and thus it is typically ignored without consequence. For ultra-narrowband filters however, this shift can easily be larger than the bandwidth of the filter itself \citep[e.g.][]{Baker19}. The success of our experiment therefore depended largely on the success of our wavelength calibration. 
 

\subsection{Wavelength Calibration with Telluric OH Lines} \label{sec:wavcal}

To begin calibrating refractive effects, we used known telluric emission lines in the sky background to construct a model for the position-dependent wavelength shift. We used a sky background frame (constructed with a four-point dither near WASP-69) shown in Figure~\ref{fig1}a. Rays that pass through the filter at the same angle of incidence trace out semi-circular arcs across the detector, and telluric OH emission lines thus appear as bright arcs on the detector. The offset center of the circles towards the top of the image is due to the aforementioned $7\degree$ tilt of the filter wheel; if the filter wheel was not tilted, the circles would be centered on the detector \citep[see e.g.][]{Sing11}. Instead, the center of the circle to which the arcs belong is the ``zero point" of the filter; i.e., where rays encounter the filter at normal incidence. The best-fitting circular arcs to the emission features give the detector position for the zero point: ($x_0, y_0$)~=~(1037,~2120), where the origin of the coordinate system is the bottom left corner of the image. The angle of incidence on the filter at detector position $(x, y)$ can be written as a function of the radial distance from the zero point $r = \sqrt{(x - x_0)^2 + (y - y_0)^2}$:
\begin{align}
    \theta(r) &= (\mathrm{pixel\ scale})(\mathrm{magnification}) r \nonumber\\
    &= (0\farcs25\mathrm{/px})\Big(\frac{5.08\ \mathrm{m}}{5.2364\times10^{-2}\ \mathrm{m}}\Big)r \nonumber\\
    &= (24\farcs3\mathrm{/px})r, \label{eq1}
\end{align}
where the magnification is calculated as the primary mirror diameter over the beam diameter. By extracting the median count value in radial steps outward from the zero point, we construct a spectrum of the sky. To convert the spectrum into more useful wavelength units, we note that the OH emission lines in the image can be individually identified as $Q$ and $R$ branch lines from the $\nu = 5-2$ band for ground state (X$^2\Pi$) OH \citep{Bernath09, Oliva15}. Using the known wavelengths of these lines, we can fit to the equation for wavelength shift as a function of angle of incidence $\theta$ \citep[e.g.][]{Ghinassi02}:
\begin{equation}
    \lambda(\theta) = \lambda_0\sqrt{1 - \frac{\sin^2(\theta)}{n_\mathrm{eff}^2}},
    \label{eq2}
\end{equation}
where $\lambda_0$ is the central wavelength of the filter at normal incidence, and $n_\mathrm{eff}$ is the effective index of refraction for the filter. A non-linear least-squares fit to the known wavelengths of the telluric lines gives $\lambda_0 = 1084.80$ nm and $n_\mathrm{eff} = 1.948$. Combined with Equation~(\ref{eq1}), this fully specifies the wavelength solution for every pixel on the detector as a function of the distance $r$ from ($x_0, y_0$) = (1037, 2120).
The spectrum of the sky background constructed with this transformation is given in Figure~\ref{fig1}b.

\subsection{Helium Arc Lamp Calibration} \label{sec:arclamp}
We used a helium arc lamp, which is a natural source of the He I triplet in vacuum, to confirm our wavelength solution and test our knowledge of the filter transmission profile. First, we measured the spectrum of the arc lamp and the transmission spectrum of the helium filter (back-lit by white light) using an Optical Spectrum Analyzer (OSA, ThorLabs \#OSA202C). The OSA uses Fourier transform spectroscopy to deliver laboratory spectra at high resolving power ($R\sim75,000$). We show the laboratory spectra in Figure~\ref{fig1}c, where the two-component structure of the helium feature is clear (the two lines on the red side of the triplet are blended even at this resolution). 

We then installed the helium arc lamp at the Hale 200" and used it to uniformly illuminate the region of the dome normally used for flat fields. When the helium lamp is observed through WIRC, the resultant bright arc (Figure~\ref{fig1}d) is where the filter transmission profile maximally overlaps with the triplet helium feature, so during science observations we place the target within the region delineated by this arc. In practice, we take an arc lamp calibration frame before each observation, and we move the target star to a spot with a count level within $5\%$ of the peak counts in the calibration frame. Since there is a semicircular locus on the detector that satisfies this criterion, the exact location is selected during observations to optimize the number of reference stars and avoid detector regions with many bad pixels or defects. Using the same procedure as detailed in Section~\ref{sec:wavcal}, we extract the spectrum from the image in Figure~\ref{fig1}d, and use the wavelength solution from Equation~(\ref{eq2}) to convert from AOI to nm. The resulting spectrum (Figure~\ref{fig1}e) peaks at 1083.3 nm, indicating that our empirical wavelength solution correctly predicts the location of the helium triplet as measured by the lamp observation. Finally, as a test of the filter transmission profile, we convolve the laboratory measurements of the helium feature and the filter transmission profile, and overplot the result on the WIRC spectrum in Figure~\ref{fig1}e. The laboratory measurements (dot-dashed blue curve) and observations (black curve) show very good agreement.

\section{Observations} \label{sec:obs}
\subsection{Data Collection} \label{sec:coll}
We observed WASP-69b through our helium filter and beam-shaping diffuser on August 16, 2019 (UT), and we observed WASP-52b with the same setup on September 17, 2019 (UT). Before beginning both science observations, we constructed a sky background frame with a simple four-point dither. Images in the dither sequence were first sigma-clipped to remove the sources, then median scaled to the first image in the stack, and finally median stacked to produce the sky background frame. We then collected science data, choosing exposure times to keep the maximum count level for the sources and comparison stars ($\sim12,000$~ADU) well within the linearity regime for our detector while maintaining a good observing efficiency. For WASP-69b, we collected science data from UT~04:26:06 to 11:00:00 with an exposure time of 60 seconds; our observations began at airmass 1.73, reached a minimum airmass of 1.28, and then rose again until we stopped collecting data at airmass 2.49. For WASP-52b, we collected science data from UT~03:16:57 to 11:14:49 with an exposure time of 90 seconds; our observations began at airmass 2.04, reached a minimum airmass of 1.10, and then rose again until we stopped collecting data at airmass 1.96. 

\subsection{Data Reduction} \label{sec:reduc}
\begin{figure*}[p!]
    \centering
    \includegraphics[width=\textwidth]{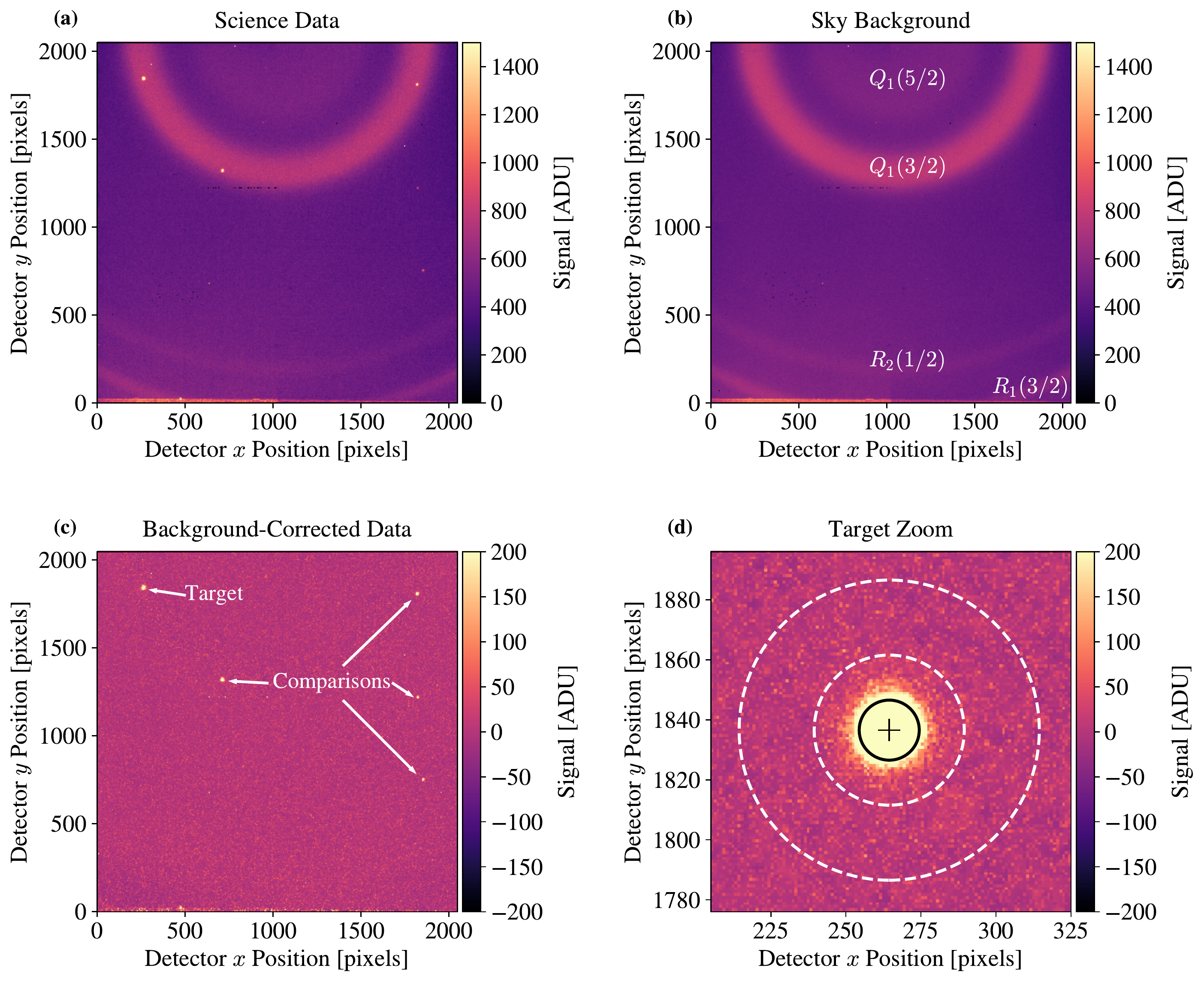}
    \caption{Example of the data reduction process. (a) A calibrated science frame from the WASP-69 observations before background correction. (b) The dithered sky background frame, with telluric lines indicated (see also Figure~\ref{fig1}). (c) The background-corrected science frame, with target and comparison stars marked. (d) A zoomed image of the target star in the background-corrected science frame (with flux-weighted centroid given by the black cross, the optimized 10 pixel aperture by the black circle, and the annulus used for residual background estimation by the white dashed circles).}
    \label{fig2}
\end{figure*}

\subsubsection{Image Calibration} \label{sec:calib}
We show an example science image for WASP-69b in Figure~\ref{fig2}. All science data were dark-subtracted and flat-fielded, and during this procedure bad pixels were flagged and corrected using the process described by \citet{Tinyanont19} and \citet{Vissapragada20}. Unlike the case in \citet{Vissapragada20}, however, the background is not uniform across the detector. Contamination from telluric OH emission is clearly visible, but because these lines have a very unique spatial structure their contribution can be identified and removed. Presently, we do not correct for telluric water during image calibration. We note that the water line at 1083.507~nm (vacuum wavelength in the observer rest frame) can potentially affect the observations, though it is diluted by a minimum of $\sim$20\% by the filter transmission at the target position. This line does not encroach upon the helium triplet unless the triplet is redshifted by $48.7~\mathrm{km/s}<v<83.6~\mathrm{km/s}$ relative to the observer. This does not occur for WASP-69b and WASP-52b (and in fact we do not observe targets at such velocity shifts because the helium signal would be spatially shifted from the positions set by the calibration lamp) so our measurements are not directly biased by telluric water. Variations in the water column, however, may indirectly affect observations by manifesting as additional noise in our light curves. Due to the narrow width of the water line \citep[$\sim0.03$~nm FWHM][]{Allart18, Nortmann18, Salz18, Allart19, AlonsoFloriano19}, relative to the filter (0.635~nm FWHM), variations would need to be large ($\sim$10\%) on timescales comparable to our exposure times ($\sim$1~min) to manifest above the photometric noise as extra white noise. Smaller variations over long timescales could manifest as a time-correlated trend in our photometric data. If warranted by the data in the future, we could correct such time-correlated variations with a Gaussian process, but we see little evidence of this effect in our final light curves. 

To correct for telluric OH emission in each image, we median-scaled our sky background frame to the sigma-clipped science data in 10 pixel steps radially outwards from the filter zero point (where, as in Equation~\ref{eq1} above, the pixel scale is 0\farcs25). This procedure removed a majority of the telluric background as shown in Figure~\ref{fig2}, but in some images left a small amount of residual local structure with maximum amplitude of 10~ADU/pixel, perhaps due to spatial variation of OH emission on the sky. Because even these residuals were locally quite stable, we estimated and removed the remaining background during aperture photometry using an annular region around each source as described below. This local background varies quite slowly in time and we find that this procedure reliably eliminates time-correlated noise from sky background and tellurics.

\subsubsection{Aperture Photometry} \label{sec:aperphot}
We detected and registered the positions of the target and comparison stars using Aladin Lite \citep{Bonnarel00, Boch14} as described in \citet{Vissapragada20}. For both WASP-69 and WASP-52, we registered four comparison stars in addition to the target; for WASP-69, the target and comparison stars are visible in the background-corrected image in Figure~\ref{fig2}. We performed aperture photometry on each source in each image with the \texttt{photutils} package \citep{Bradley16} where we stepped through a range of circular apertures (from 7 to 15 pixels in radius in one pixel steps). The positions of the aperture centers were allowed to shift to trace telescope pointing drift. For WASP-69 and associated comparison stars, these varied by less than 2 pixels over most of the night, but a guiding error compromised the last hour of data collection. Excluding this last hour did not change our final answers but substantially decreased the correlated noise, so we choose to exclude these images from the final photometry. For WASP-52 we encountered a guiding jump of about 6 pixels an hour from the start of the observation, and again an hour from the end of the observation. These jumps were purely in the RA direction and are thus likely related to a known issue with the RA guiding on the telescope. Including the data marred by guiding errors substantially increased the correlated noise in the final light curve, so we opted to leave them out for our analysis of WASP-52b.

\begin{figure*}[ht!]
    \centering
    \gridline{\fig{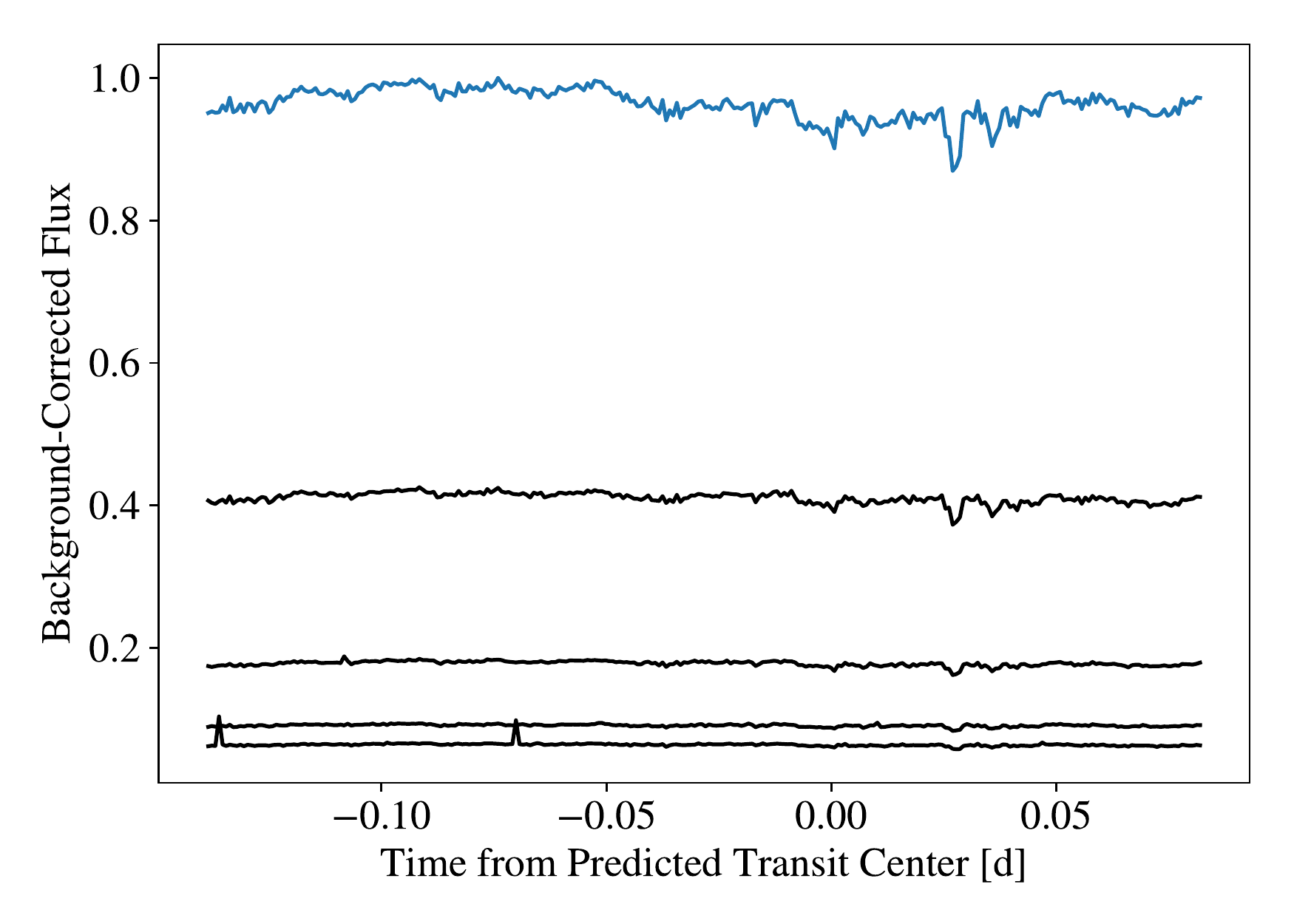}{0.47\textwidth}{(a)}
          \fig{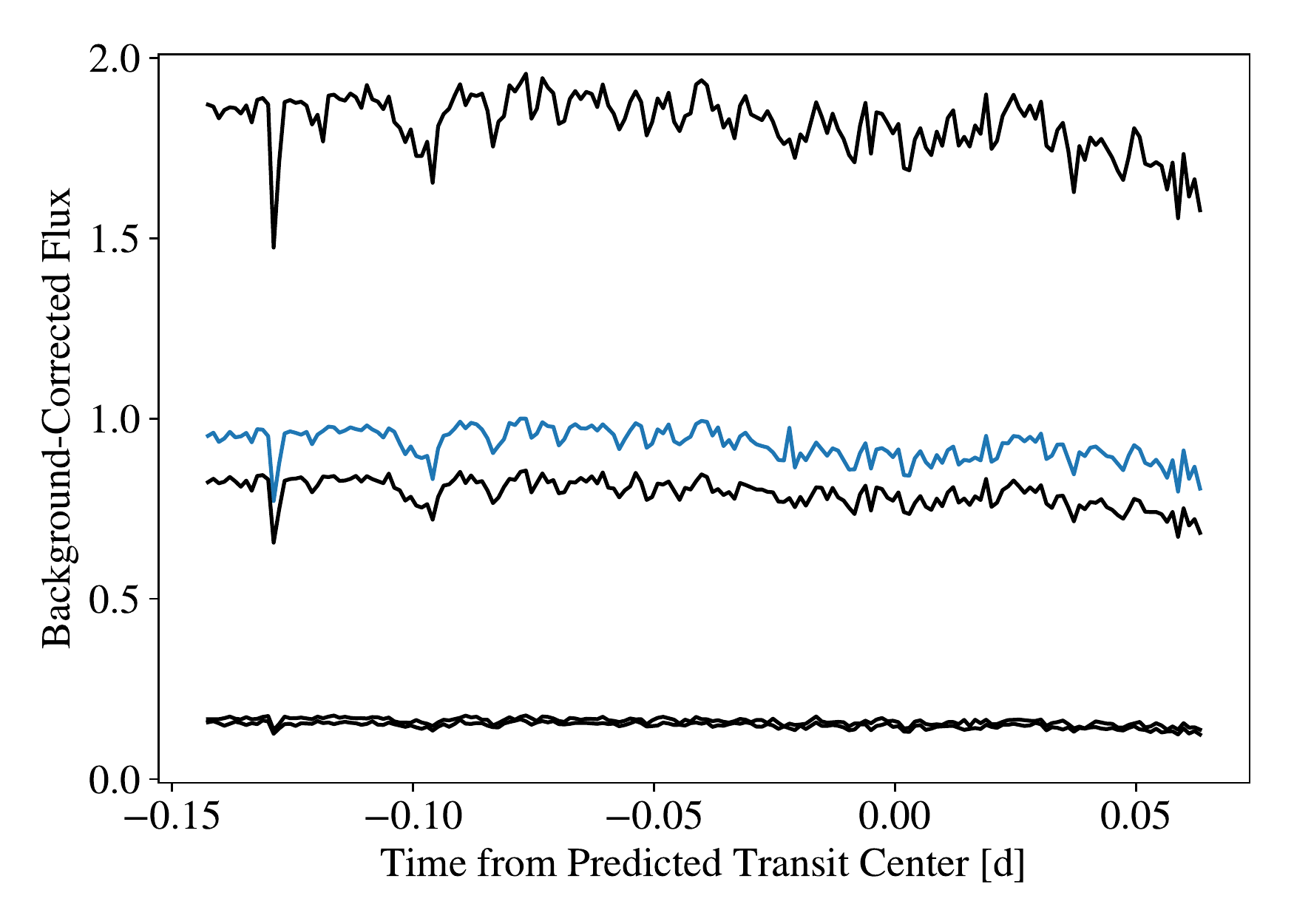}{0.47\textwidth}{(b)}}
          
    \caption{Raw light curves for stars in the WASP-69 field (a) and WASP-52 field (b). In both plots, the target light curve is shown in blue, comparison light curves are shown in black, and all light curves have been normalized to the target light curve maximum.}
    \label{fig:rawlc}
\end{figure*}

We estimated the residual local background by measuring the sigma-clipped median for an annulus around each source with an inner radius of 25 pixels and an outer radius of 50 pixels. We then trimmed outliers in the raw light curves using the moving median procedure from \citet{Vissapragada20}. We determined the optimal photometric aperture size by minimizing the RMS of the residuals after the light-curve modeling described in the next section.  Our optimal apertures were 10 and 8 pixels in radius for WASP-69 and WASP-52, respectively. A zoomed-in view of WASP-69 with flux-weighted centroid, best-aperture, and background annulus overplotted is shown in Figure~\ref{fig2}. It is clear from this figure that a 10 pixel aperture misses some flux from the target star. However, when the aperture size increases to encompass all of the flux from the target star, the comparison star light curves decrease in quality due to increased noise from the sky background. We tested the impact of using different aperture sizes for each source and found that this sharply degraded the quality of the final light curve, likely because PSF changes due to seeing variations impact each aperture differently. We therefore chose to continue with the selected optimal apertures in our final light-curve modeling. Raw light curves in the optimized apertures are given in Figure~\ref{fig:rawlc} for both planets.

\subsection{Light-Curve Modeling}
We modeled the light curves with a procedure similar to that used in \citet{Vissapragada20}, which we briefly summarize here for completeness. Each target light curve is modeled as a transit light-curve model \citep[which is computed with \texttt{batman};][]{Kreidberg15} multiplied by a systematics model. The systematics are further modeled as a linear trend in time plus a linear combination of the comparison star light curves, with new best-fitting linear coefficients chosen every time the transit light curve is modified. As in \citet{Vissapragada20}, our six fit parameters were the transit depth $(R_\mathrm{p}/R_\star)^2$, a timing offset from the predicted mid-transit time $\Delta t_0$, a linear trend in time $\alpha$, the inclination $i$, the scaled semi-major axis $a/R_\star$, and a parameter describing the photometric scatter in excess of shot noise $\log(\sigma_\mathrm{extra})$. The excess scatter that we calculate is added in quadrature to the photometric error bars on each data point to give the final errors. We calculated custom quadratic limb darkening coefficients $u_1$ and $u_2$ in our bandpass using \texttt{ldtk} \citep{Husser13, Parviainen15} and the stellar parameters from \citet{Anderson14} and \citet{Hebrard13} for WASP-69 and WASP-52, respectively. These coefficients are reported in Table~\ref{table1}. We additionally explored the possibility of fitting the quadratic limb darkening coefficients using the triangular sampling algorithm from \citet{Kipping13}, but found that this did not make a substantive difference in our final results, so we chose to leave these coefficients fixed.

We first fit the data using the Powell minimizer from \texttt{scipy} \citep{Jones01}, and we use this initial solution as a starting point for a Markov Chain Monte Carlo investigation with \texttt{emcee} \citep{ForemanMackey13}. We run 50 chains for $10^3$ steps to burn in, and then $10^4$ steps (which corresponds to at least 150 integrated autocorrelation times for each parameter) for the actual run. The posteriors from these light-curve fits are summarized in Table~\ref{table1}, and they are visualized in Appendix~\ref{ap:posteriors}. 

  \begin{deluxetable*}{cccccc}
  \tablecolumns{6} \tablewidth{900pt}
 \tablecaption{Light-Curve Fitting Results}
   \tablehead{\colhead{Parameter} & \multicolumn{2}{c}{{Prior}} & \multicolumn{2}{c}{Posterior} & Note \\ & \colhead{WASP-69b} & \colhead{WASP-52b} & \colhead{WASP-69b} & \colhead{WASP-52b}}
  \startdata
    $P$ (days) & 3.86814098 & 1.74978179 & (fixed) & (fixed) & (1), (2)\\
    $t_0$ (BJD$_\mathrm{TDB}$) & 2458711.8300727 & 2458743.8135163 & (fixed) & (fixed) & (1), (2)\\
    $u_1$ & 0.3975 & 0.3635 & (fixed) & (fixed) & (3), (4), (5) \\
    $u_2$ & 0.1156 & 0.1229 & (fixed) & (fixed) & (3), (4), (5) \\
    $e$ & 0. & 0. & (fixed) & (fixed) & (4), (5) \\
    \tableline
    $(R_\mathrm{p}/R_\star)^2$ (\%)& $\mathcal{U}(0.0, 3.0)$  & $\mathcal{U}(0.0, 6.0)$ & 2.152$^{+0.045}_{-0.045}$ & 2.97$^{+0.13}_{-0.13}$ &  -- \\
    $\Delta t_0$ (min) & $\mathcal{N}(0.0, 0.70)$ & $\mathcal{N}(0.0, 0.65)$ & -0.57$^{+0.42}_{-0.42}$ & -0.39$^{+0.54}_{-0.54}$ & (1), (2)\\
    $i$ ($\degree$) & $\mathcal{N}(86.71, 0.20)$ & $\mathcal{N}(85.17, 0.13)$ & 86.63$^{+0.15}_{-0.15}$ & 85.20$^{+0.12}_{-0.12}$ & (4), (6) \\
    $a/R_\star$ & $\mathcal{N}(12.00, 0.46)$ & $\mathcal{N}(7.22, 0.07)$ & 11.82$^{+0.25}_{-0.25}$ & 7.207$^{+0.062}_{-0.062}$ & (4), (6)\\
    $\alpha$ & $\mathcal{U}(-0.2, 0.2)$ & $\mathcal{U}(-0.2, 0.2)$ &   0.0160$^{+0.0026}_{-0.0025}$ & 0.0811$^{+0.0012}_{-0.0012}$ & --\\
    $\log(\sigma_\mathrm{extra})$ & $\mathcal{U}(-3.5, -2.0)$ & $\mathcal{U}(-3.5, -2.0)$ & -2.711$^{+0.025}_{-0.025}$ & -2.422$^{+0.060}_{-0.070}$ & -- \\
   \enddata
\tablecomments{(1) WASP-69b ephemerides from \citet{Basturk19}; (2) WASP-52b ephemerides from \citet{Baluev19}; (3) Quadratic limb darkening coefficients calculated with \texttt{ldtk} \citep{Husser13, Parviainen15} (4) Stellar parameters (for limb darkening calculations), $e$, $i$, and $a/R_\star$ from \citet{Anderson14} for WASP-69b; (5) Stellar parameters (for limb darkening calculations) and $e$ from \citet{Hebrard13} for WASP-52b; (6) $i$ and $a/R_\star$ from \citet{Alam18} for WASP-52b. Note also that $\mathcal{N}(a, b)$ denotes a Gaussian distribution centered on $a$ with standard deviation $b$, and $\mathcal{U}(a, b)$ denotes a uniform distribution between $a$ and $b$.}
\end{deluxetable*}
  \label{table1}

\label{sec:model}
\section{Results and Discussion} \label{sec:res}

\begin{figure*}[p!]
    \centering
    \gridline{\fig{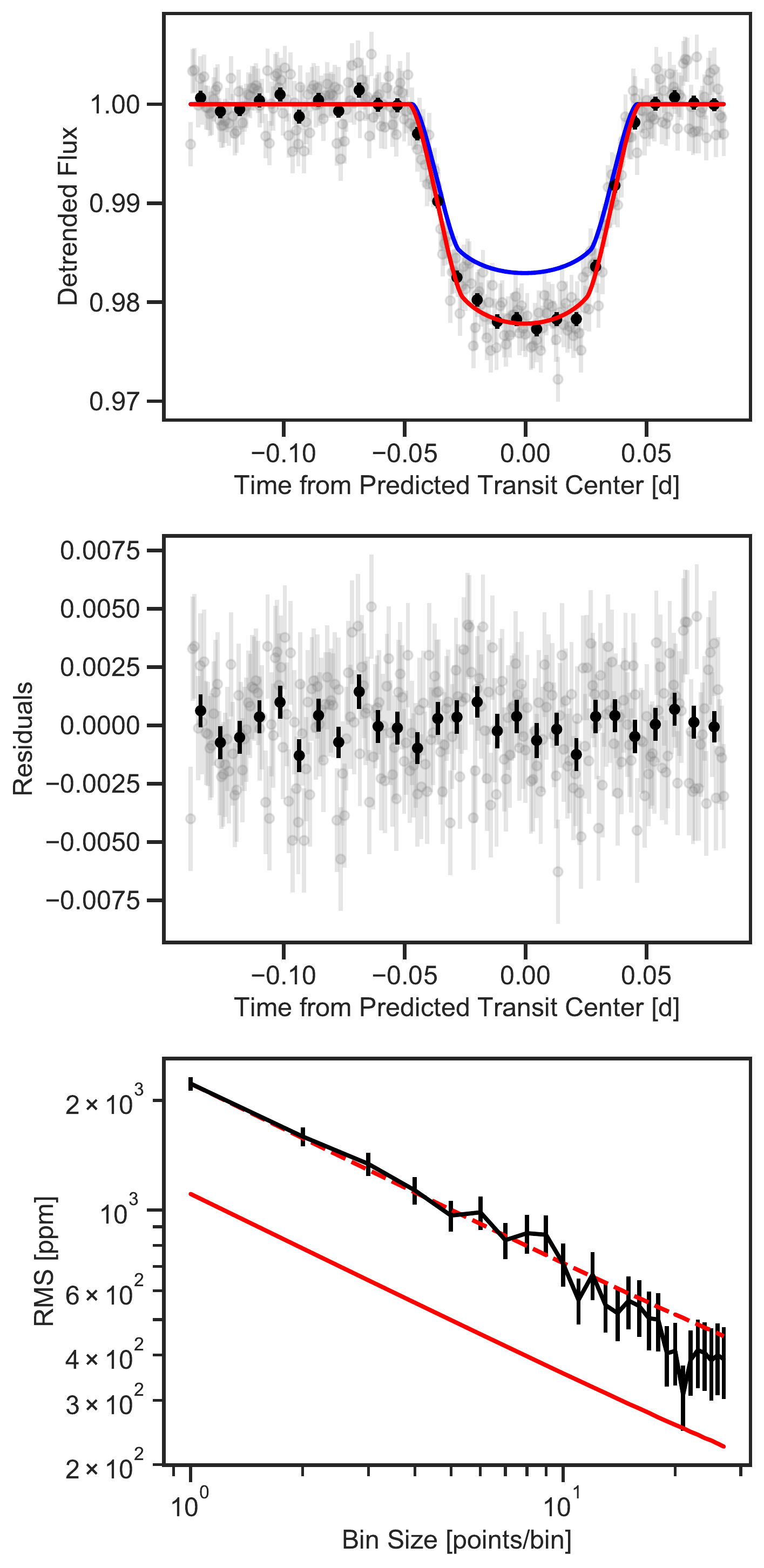}{0.47\textwidth}{(a)}
          \fig{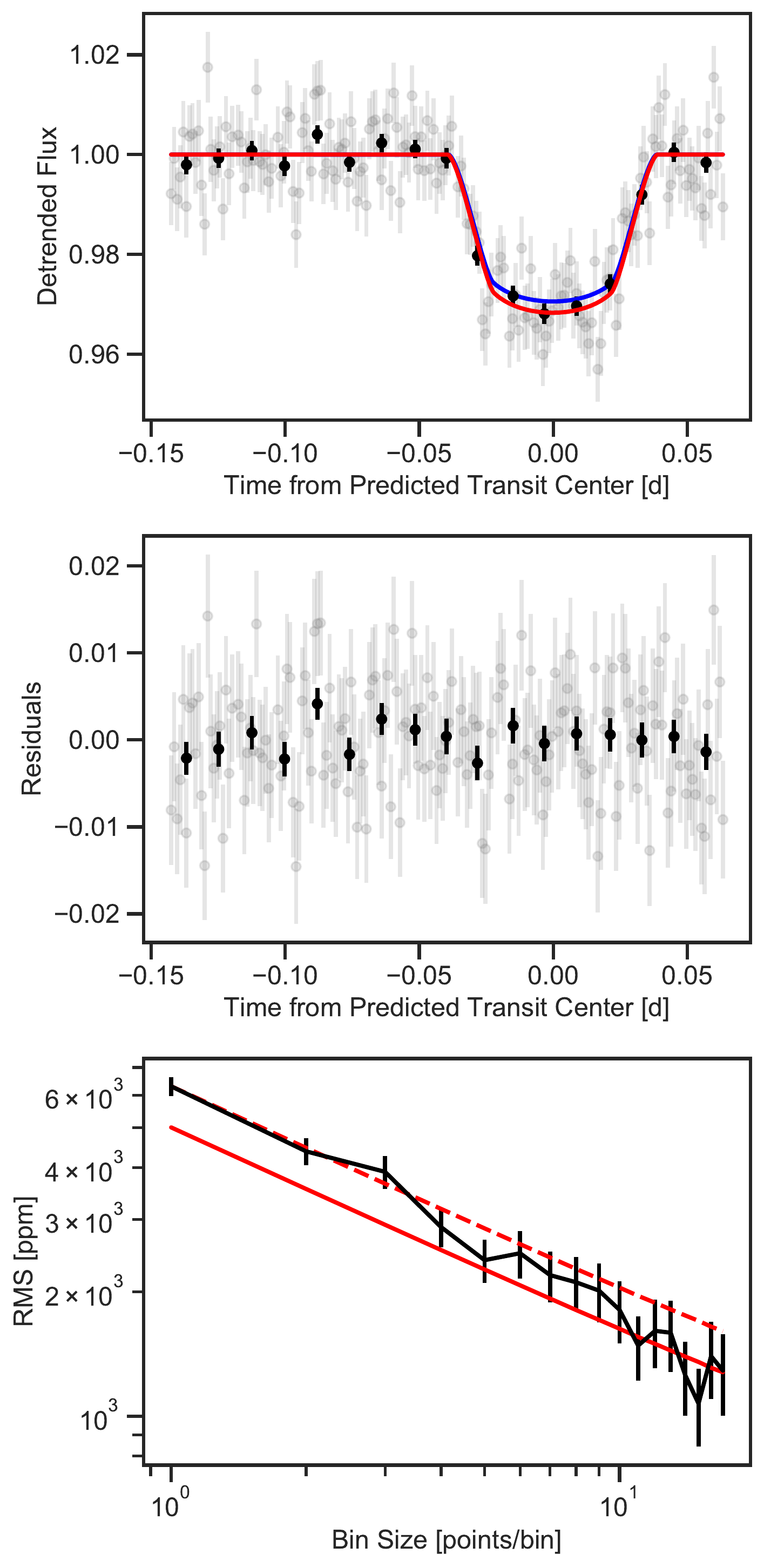}{0.47\textwidth}{(b)}}

    \caption{Results for WASP-69b in (a) and WASP-52b in (b). Top: helium light curves, with unbinned data in gray and data binned to a 10 minute cadence in black, with best-fit models shown by the red curves. The blue curves indicate reference transit depths from \citet{Tsiaras18} for WASP-69b and \citet{Alam18} for WASP-52b. Middle: fit residuals, with unbinned data in gray and binned to 10 minute cadence in black. Bottom: Allan deviation plot of the residuals (black curve) along with the photon noise limit (red curve) and the predicted behavior of our residuals assuming white noise statistics (red dashed line). We find that the scatter in these data is 2.0$\times$ the photon noise limit for WASP-69b and 1.3$\times$ the photon noise limit for WASP-52b.}
    \label{fig3}
\end{figure*}

\subsection{WASP-69b}
Our helium light curve for WASP-69b, along with best-fit model, residuals, and Allan deviation plot for the residuals are shown in Figure~\ref{fig3}a, and a corner plot summarizing the fit posteriors is shown in Figure~\ref{fig6}. We measure a transit depth of 2.152$\pm0.045\%$. As a reference value, we use the \textit{HST} WFC3 spectrum obtained by \citet{Tsiaras18}, who report an average transit depth of $1.6538\pm 0.0045\%$ between 1110.8 nm and 1141.6~nm. Our transit depth exceeds the reference value by 11.1$\sigma$, indicating a secure detection of helium in the atmosphere of WASP-69b. We prefer a transit timing solution slightly earlier than, but not incompatible with, the ephemeris from \citet{Basturk19}. Our constraints on $i$ and $a/R_\star$ are compatible with those from \citet{Anderson14}. We note, however, slight covariances between these parameters and the transit depth in Figure~\ref{fig6}. Updated knowledge on these parameters may allow us to better constrain the transit depth in the future. 

We achieved a per-point rms of 8.21 ppm/pt across 271 points. The final scatter in our residuals was 2.0$\times$ the shot noise (the noise floor set by Poisson statistics on our total detected photon counts, of which approximately 25\% are background counts due to OH emission). A small correlated component to the noise appears on 10~minute timescales (see Figure~\ref{fig3}a); we obtain a \citet{Carter09} $\beta$ factor of 1.08. This is noticeably larger scatter (relative to shot noise) than what we have typically achieved in the past for targets of similar apparent brightness \citep{Vissapragada20}. We observed this target at high efficiency (collecting light 87.6\% of the time we were on sky), and the long exposure times make scintillation noise an unlikely culprit \citep{Stefansson17}. This may be a signature of variation in the stellar He I line itself \citep{Sanz-Forcada08, Andretta17, Salz18}, but if such variations occur on long timescales (e.g. from spots on the stellar surface), then they would be corrected by our linear detrending model, and if they occur on short timescales, they would manifest as strong red noise in the light curve, which we do not observe.
Rather, the likely explanation for our photometric performance is a paucity of good comparison stars in the field. WASP-69 inhabits a fairly sparse field already, and to compound the issue we are limited in target placement to the arc shown in Figure~\ref{fig1}d, which may put otherwise accessible comparison stars outside the field of view. Thus, we are limited in our ability to obtain many good comparison stars for this technique, which here is likely the ultimate limiting factor in our photometry.

\begin{figure*}[ht!]
    \centering
    \gridline{\fig{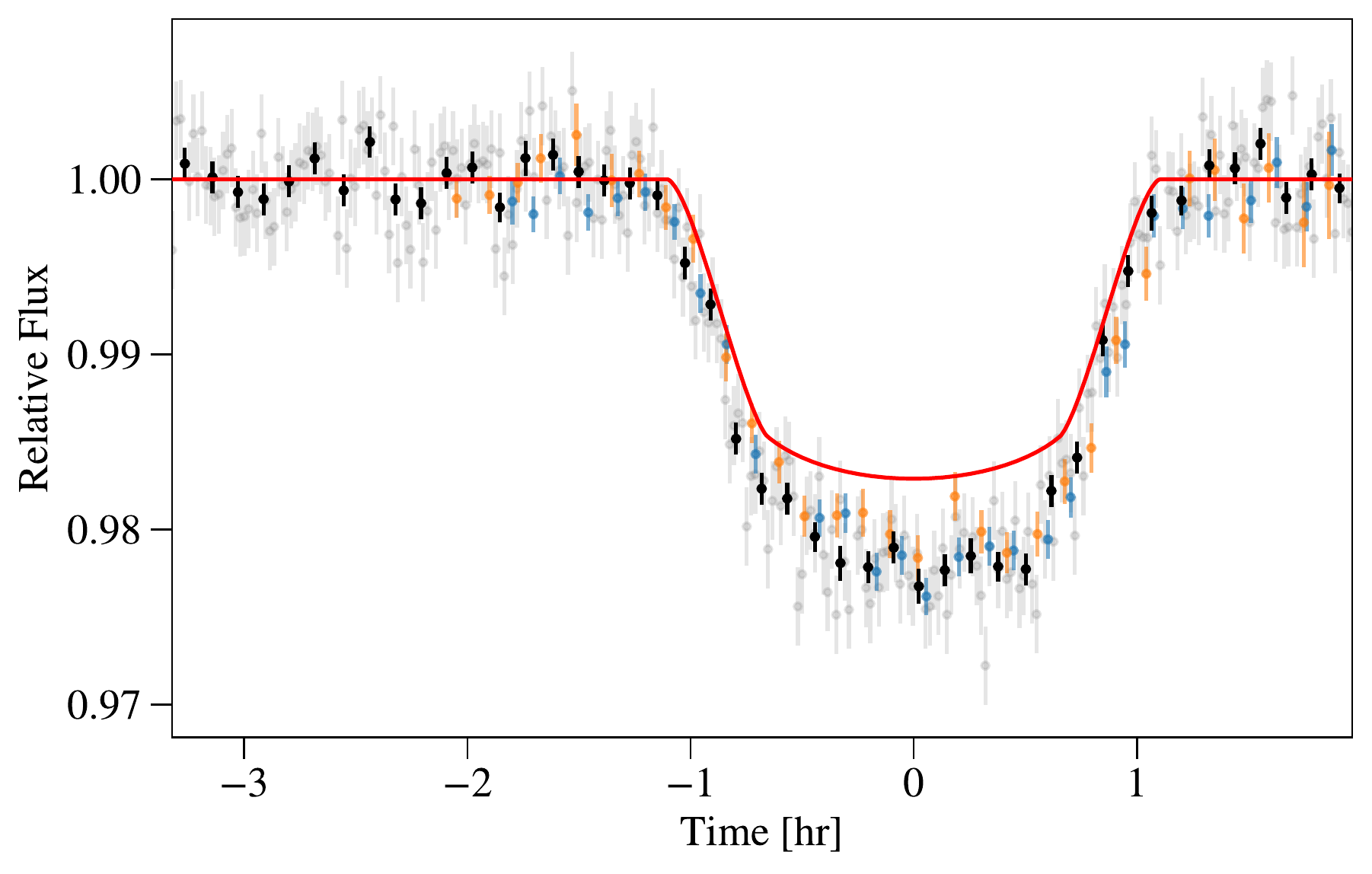}{0.47\textwidth}{(a)}
          \fig{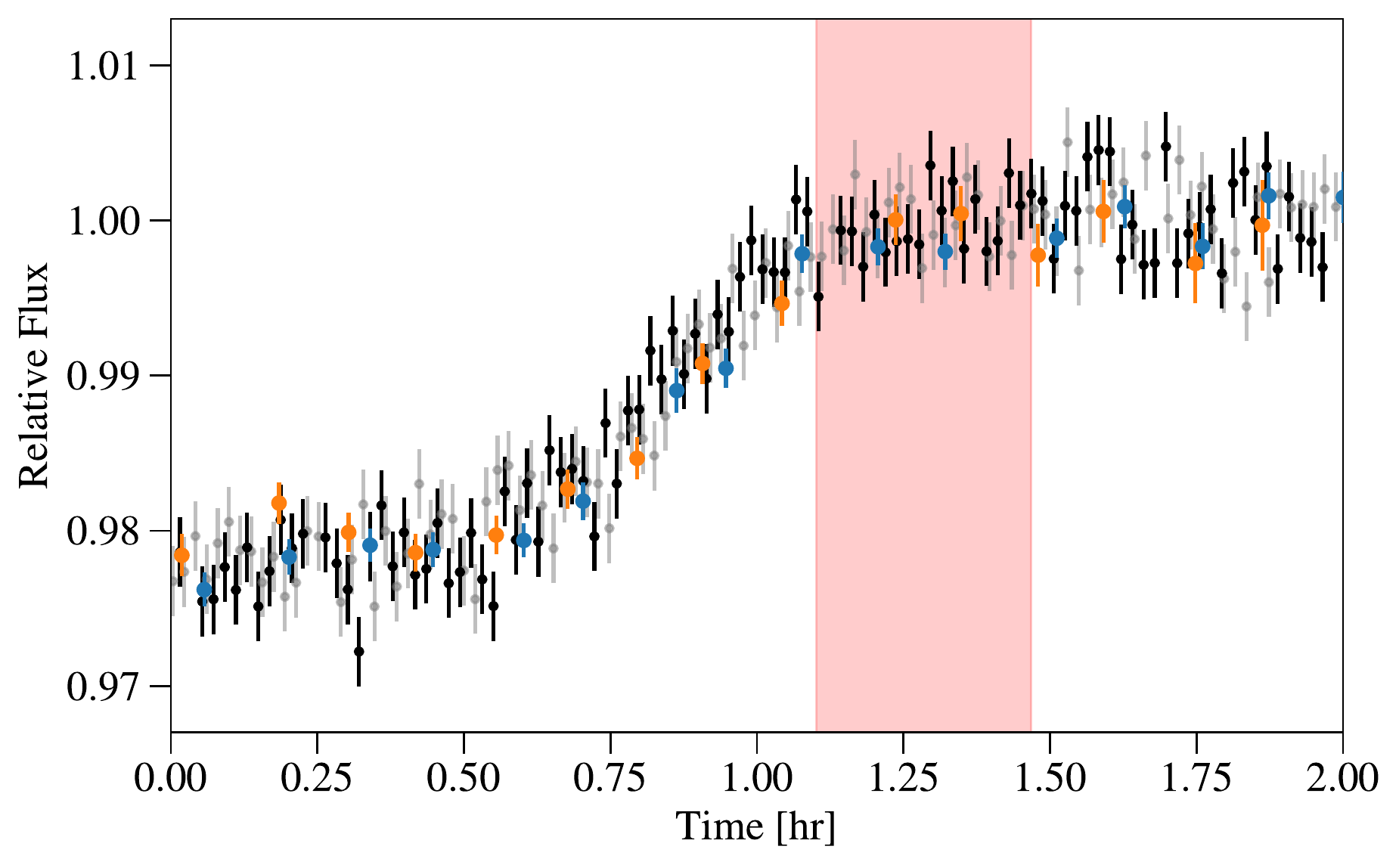}{0.47\textwidth}{(b)}}

    \caption{(a) WIRC light curve of WASP-69b (unbinned in gray and binned to 7 minute cadence in black) compared to CARMENES light curves (computed by integrating CARMENES spectra against our transmission function) from \citet{Nortmann18} in blue and orange (their first and second nights of data collections, respectively). The comparison light curve from \citet{Tsiaras18} is shown in red. (b) Mirrored, unbinned WIRC light curve, with ingress shown in gray and egress shown in black. Data from CARMENES are again shown in blue and orange for the first and second nights of data collection \citep{Nortmann18}. The post-egress absorption reported by \citet{Nortmann18} would fall within the red region. We do not see significant evidence for it here, but the asymmetry is also washed out in the calculated CARMENES light curve due to our wide bandpass (relative to the CARMENES resolution element).}
    \label{fig4}
\end{figure*}

We now assess how our transit measurement compares to the spectroscopic measurement of \citet{Nortmann18}. We took their reduced stellar spectra gathered over two nights of observation and converted these from the planet rest frame (in which the reduced data were provided) back to the telluric rest frame. For each spectrum (which we label $f_{i,\lambda}$, where $i$ indexes time and $\lambda$ indexes wavelength), we calculated the excess absorption signal $f_i$ in our bandpass using our measured transmission function $T_\lambda$ via
\begin{equation}
    f_i = \frac{\int f_{i,\lambda} T_\lambda d\lambda}{\int T_\lambda d\lambda}.
\end{equation}
The timeseries $f$ then represents the excess absorption in the helium line during the transit as would be measured by CARMENES through our helium filter. To this we added the broadband light curve \citep[calculated with the parameters of][]{Tsiaras18} which gave the total light curve as would have been observed by WIRC. We repeated this procedure for both nights of CARMENES data collection (with 35 spectra in night 1 and 31 spectra in night 2), and we present our results compared to the two CARMENES timeseries in Figure~\ref{fig4}a. Our data show good agreement with those collected by \citet{Nortmann18}. 

\citet{Nortmann18} also report the detection of an asymmetric transit in He I, with egress extending about half an hour past ingress. We do not find strong evidence for this effect in our light curve. In Figure~\ref{fig4}b, we show our WASP-69b light curve mirrored across our best-fit mid-transit time; there is no visible absorption in the post-egress window where \citet{Nortmann18} report an extended tail. While we do not see strong evidence for this effect in our light curve, however, we cannot rule it out. The amplitude of the reported post-egress absorption is of order $0.5\%$; when diluted through our transmission function this becomes a $500$~ppm effect which we are not significantly sensitive to on a 22~min timescale (our rms on this timescale is 388 ppm). Repeated observations of WASP-69b may allow us to constrain the transit asymmetry in the future.

\subsection{WASP-52b}
Our helium light curve for WASP-52b, along with best-fit model, residuals, and Allan deviation plot for the residuals are shown in Figure~\ref{fig3}b, and a corner plot summarizing the fit posteriors is shown in Figure~\ref{fig7}. We measure a transit depth of 2.97$^{+0.13}_{-0.13}$\%, which exceeds the spot-uncorrected transit depth between 898.5 nm and 1030.0 nm ($2.76\pm0.021\%$) from \citet{Alam18} by 1.6$\sigma$. Assuming the same line structure shape as is observed for WASP-69b \citep{Nortmann18}, this converts to an amplitude of $1.31\pm0.94\%$ in the deepest line of the triplet. This is meant only to give a sense of what one might expect at high resolution; in reality, lineshapes can vary from planet to planet, and there is no guarantee that assuming the line shape of WASP-69b is correct \citep{Nortmann18, Allart18, Salz18, Allart19, AlonsoFloriano19, Kirk20}. We obtained a per-point RMS of 35.6 ppm/pt across 177 points. The scatter in the light curve was 1.3$\times$ the photon noise limit, binning down like white noise (see bottom panel of Figure~\ref{fig3}). This performance is comparable to what we have achieved in the past for similar targets \citep{Vissapragada20}, despite the fact that there were only four comparison stars in the field of view.

WASP-52 is a young ($0.4^{+0.3}_{-0.2}$~Gyr), active host star, with a $\log R'_\mathrm{HK}$ index of $-4.4\pm0.2$ \citep{Hebrard13} and many authors observing and analyzing the effects of spots and plages \citep{Kirk16, Chen17, Louden17, Mancini17, Alam18, Bruno18, May18, Bruno19}. Considering the proposed relationship between planetary metastable helium absorption and stellar activity \citep{Nortmann18, AlonsoFloriano19}, WASP-52 remains a high-priority target for future work. Follow-up observations with high-resolution spectroscopic facilities on larger telescopes should be able to detect absorption and quantify the line shape (which we must assume here) for this rather challenging target. We note that confident detections of Na, K, and H$\alpha$ absorption in the atmosphere of this planet recently required three transits with the ESPRESSO high-resolution spectrograph on the VLT \citep{Chen20}. Though its host star is relatively faint, WASP-52b is well worth additional observations in metastable helium, as the other detected atomic species will provide some context for modeling the upper atmosphere of this planet.

\subsection{Mass Loss Modeling}

\begin{figure*}[ht!]
    \centering
    \gridline{\fig{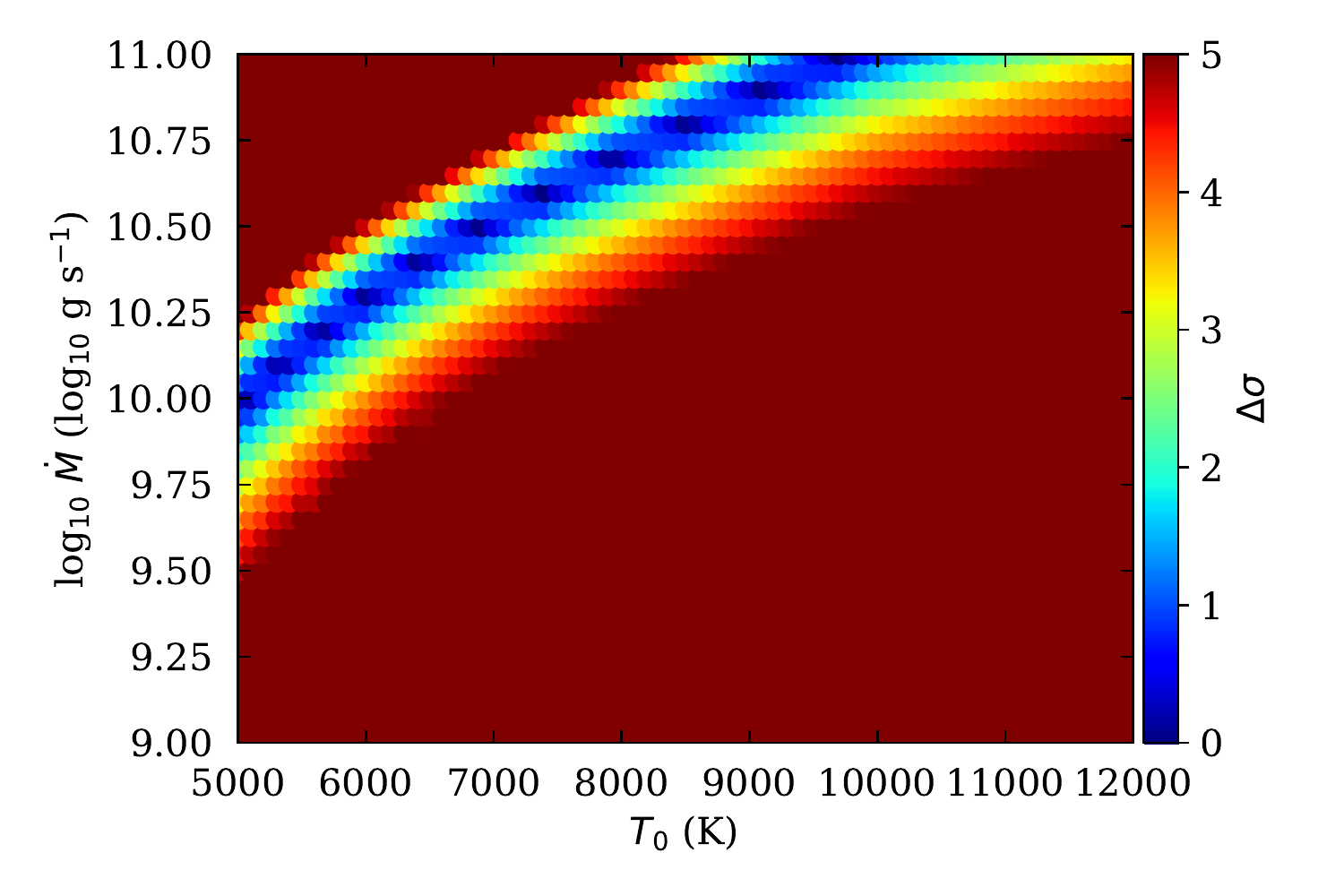}{0.47\textwidth}{(a)}
          \fig{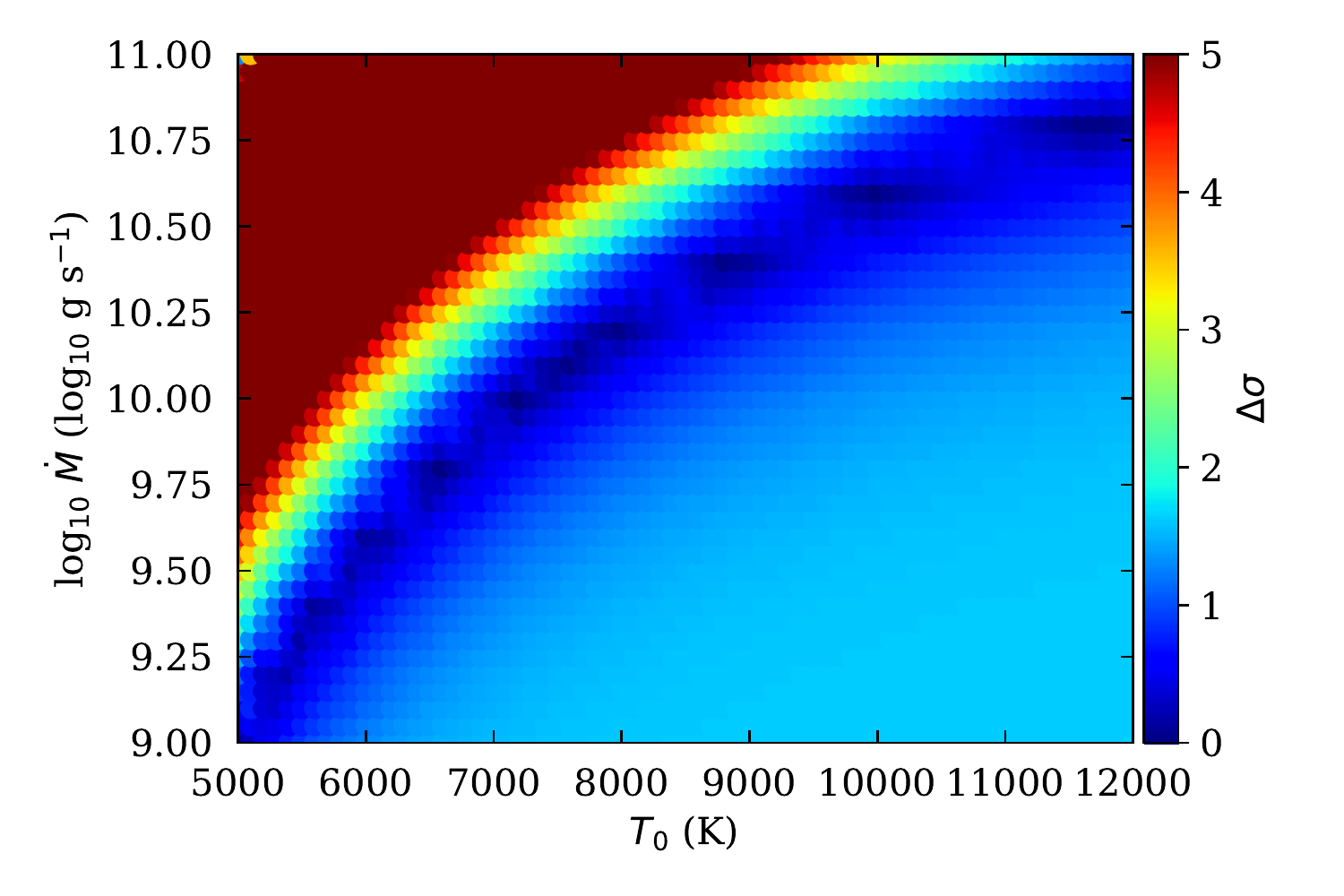}{0.47\textwidth}{(b)}}
    \caption{Mass loss modeling for WASP-69b in (a) and WASP-52b in (b). Each point $(T_0, \dot{M}$) corresponds to a different mass loss model, and the color of the point indicates the $\sigma$ discrepancy between that model and the data presented in Figure~\ref{fig4}.}
    \label{fig5}
\end{figure*}

We interpret our observations of WASP-69b and WASP-52b using the \citet{Oklopcic18} model. Despite our lack of a significant detection for WASP-52b, we model potential outflows from this planet to set an upper limit on the mass loss rate corresponding to our upper limit on the excess absorption. As WASP-52b is a high-priority target for future observations \citep{Kirk20}, this is a particularly important constraint that we can obtain from our light curve.

We first computed grids of atmospheric mass loss models; following \citet{Oklopcic18} and \citet{Mansfield18}, we computed 1D density and velocity profiles for a 90\%--10\% hydrogen--helium atmosphere losing mass to an isothermal Parker wind. These profiles spanned 5,000--12,000~K in thermosphere temperature $T_0$ and $10^9$--$10^{11}$~g/s in mass loss rate $\dot{M}$, with the ranges motivated by hydrodynamics simulations of atmospheric escape \citep{Salz16}. Level populations for hydrogen and helium were then computed for each profile. As there are no measurements of the stellar UV spectra (required for computing photoionization rates) for WASP-69 and WASP-52, we used UV spectra from MUSCLES \citep{France16} of stars with similar spectral type. For WASP-69, we used HD 85512 (K6) and for WASP-52 we used  $\epsilon$ Eri (K2). 

The resulting density profiles of 2$^3$S He were then used to compute the transit depth in the line given our filter transmission function, and the model transit depths were compared to those that we report in Table~\ref{table1}. We opted to compare only the transit depths from the outflow models to our data rather than the full light curve, as the full computation is substantially more expensive for a marginal gain in accuracy for the model comparison (relative to our photometric uncertainties). In Figure~\ref{fig5}, we show how the model grids compare to our data, parameterized by the number of standard deviations away from our data. For WASP-69b we obtain a curved contour of best-fit solutions, indicating a known degeneracy between mass loss rate and thermosphere temperature due to our inability to resolve line shapes \citep{Mansfield18}. 

To summarize the contours in Figure~\ref{fig5}, we quote our constraints on the mass loss rate at two possible thermosphere temperatures.  At $T_0 = 7,000$~K (12,000~K) we obtain a corresponding mass loss rate of $\dot{M} = 10^{10.50^{+0.05}_{-0.04}}~\mathrm{g/s}$ ($\dot{M} = 10^{11.30^{+0.08}_{-0.08}}~\mathrm{g/s}$). This translates to $5.25^{+0.65}_{-0.46}\times10^{-4}~M_\mathrm{J}/\mathrm{Gyr}$ ($3.32^{+0.67}_{-0.56}\times10^{-3}~M_\mathrm{J}/\mathrm{Gyr}$). The mass loss rate for WASP-69b is therefore very similar to those reported for HAT-P-11b and WASP-107b \citep{Allart18, Mansfield18, Spake18, Allart19, Kirk20}, which should be typical for planets at similar distances and gravitational potentials \citep{Salz16}. For WASP-52b, we can set a 95th-percentile upper limit of $\dot{M}<10^{10.1}~\mathrm{g/s}\ (10^{11.1}~\mathrm{g/s})$ at $T_0 = 7,000$~K (12,000~K). This translates to $2.1\times10^{-4}~M_\mathrm{J}/\mathrm{Gyr}\ (2.1\times10^{-3}~M_\mathrm{J}/\mathrm{Gyr})$. We conclude from these measurements that, barring substantial changes in orbital distance and stellar irradiation, WASP-69b ($M_\mathrm{p} = 0.26M_\mathrm{J}$) and WASP-52b ($M_\mathrm{p} = 0.46M_\mathrm{J}$) will survive over the lifetime of their host stars (losing at most a few percent in envelope mass), and their compositions will not be substantially impacted by mass loss.



\section{Conclusions} \label{sec:conc}
In this work, we have presented a new photometric technique to observe the metastable 2$^3$S helium absorption feature near 1083.3~nm~using an ultra-narrowband filter and a beam-shaping diffuser. We benchmarked this new technique by observing WASP-69b, a planet for which the shape of the helium feature has been measured with high-resolution spectroscopy \citep{Nortmann18}. Our technique detects helium absorption to 11.1$\sigma$ confidence (a single-transit S/N comparable to that achieved with CARMENES) in this planet's atmosphere, at a level consistent with previous observations. Additionally, for WASP-52b we set a 95th-percentile upper limit on excess absorption in the helium bandpass of 0.47\%. We find that the quality of our photometry relative to the photon noise limit depends sensitively on the availability of comparison sources. Interpreting our results with atmospheric mass loss modeling allows us to constrain the mass loss rate for WASP-69b to  $5.25^{+0.65}_{-0.46}\times10^{-4}~M_\mathrm{J}/\mathrm{Gyr}$ ($3.32^{+0.67}_{-0.56}\times10^{-3}~M_\mathrm{J}/\mathrm{Gyr}$) at 7,000~K (12,000~K), and additionally we set an upper limit to the mass loss rate for WASP-52b at these temperatures of $2.1\times10^{-4}~M_\mathrm{J}/\mathrm{Gyr}$ ($2.1\times10^{-3}~M_\mathrm{J}/\mathrm{Gyr}$). These values are typical for other gaseous planets at similar gravitational potentials and orbital periods, and we conclude that both of these planets' atmospheres will not be substantially affected by mass loss for many Gyr.

Diffuser-assisted, ultra-narrowband photometry on a wide-field camera is a unique way to study exoplanet atmospheres, but it also comes with challenges. For the experimental setup detailed here, we sometimes have to settle for sub-optimal photometry on brighter targets because we are observing in sparse fields with relatively few suitable comparison stars, and also because of the constraints imposed by the AOI shift effect. Additionally, the lack of a comparison bandpass means that we must rely on high-precision infrared transit measurements taken by other groups (or simultaneous measurements with different instruments) to establish the magnitude of the excess absorption in the helium line, rather than doing so in our own experimental setup. Both of these challenges could be overcome with photometers like those presented in \citet{Baker19}, which allow for simultaneous photometry of a target star in two adjacent passbands. Though our restricted instrumental setup does not presently allow us to use this method, or other multi-color imaging methods requiring dichroics \citep[e.g.][]{Dhillon16}, we believe these are fruitful avenues for future exploration in the context of narrow atomic and molecular features.

Despite the challenges we have encountered in our constrained experimental setup with WIRC, we have demonstrated that our system is capable of measuring mass loss rates for most advantageous targets. Our technique occupies a unique niche in the current suite of approaches to metastable helium observations. First, the narrowband filter affords us better precision than space-based spectroscopy with \textit{HST} WFC3, scaling from the precisions of \citet{Spake18} and \citet{Mansfield18}. Second, while the \textit{James Webb Space Telescope} will achieve much better precision \citep{Allart18}, we can schedule and observe targets more readily on a ground-based 5~m telescope, allowing us to survey a wider range of planets. Third, the high efficiency of our technique lets us observe targets beyond the magnitude limits of high-resolution spectrographs on smaller telescopes. With future WIRC observations, we aim to characterize the fundamental relationships between mass loss, stellar activity, high-energy flux, and planetary age \citep{Nortmann18, AlonsoFloriano19, Oklopcic19, Owen19}.

\acknowledgments
We thank the referee for a very thorough review that improved the quality of this work. We are grateful for the support of the  Heising-Simons Foundation, which allowed us to purchase the narrowband filter used in this study. We thank Lisa Nortmann for providing us with the CARMENES data for WASP-69b. We recognize the Palomar Observatory staff for their support of our work, especially Paul Nied and Kajse Peffer for telescope operation and James Brugger, Greg Van Idsinga, Ernie Velador, and Brian Faull for assistance with helium lamp hardware. We also thank James Owen, Yanqin Wu, Trevor David, Ignas Snellen, Yayaati Chachan, Fei Dai, Munazza Alam, Nikolay Nikolov, Chaz Shapiro, Jennifer Milburn, Andy Boden, Roger Smith, and Keith Matthews for very useful conversations. SV is supported by an NSF Graduate Research Fellowship and the Paul \& Daisy Soros Fellowship for New Americans. HAK acknowledges support from NSF CAREER grant 1555095 and NASA Origins grant NNX14AD22G. CKH acknowledges support from the University of Maryland Department of Astronomy Honors Program, and from the Smithsonian Astrophysical Observatory REU program, which is funded in part by the National Science Foundation REU and Department of Defense ASSURE programs under NSF Grant no. AST-1852268, and by the Smithsonian Institution. AO acknowledges support by NASA through the NASA Hubble Fellowship grant HST-HF2-51443.001-A awarded by the Space Telescope Science Institute, which is operated by the Association of Universities for Research in Astronomy, Incorporated, under NASA contract NAS5-26555.
 
\facilities{Hale (WIRC), ADS, Exoplanet Archive}
\software{photutils \citep{Bradley16}, numpy \citep{vanderWalt11}, astropy \citep{Astropy13, Astropy18}, scipy \citep{Jones01}, matplotlib \citep{Hunter07}, batman \citep{Kreidberg15}, emcee \citep{ForemanMackey13}, corner \citep{ForemanMackey16}, ldtk \citep{Husser13, Parviainen15}, Aladin Lite \citep{Bonnarel00, Boch14}}

\appendix 
\section{Posterior Probability Distributions \label{ap:posteriors}}
\restartappendixnumbering
In this section, we show the posterior probability distributions for our light-curve fits to WASP-69b and WASP-52b. 
\begin{figure*}[ht!]
    \centering
    \includegraphics[width=\textwidth]{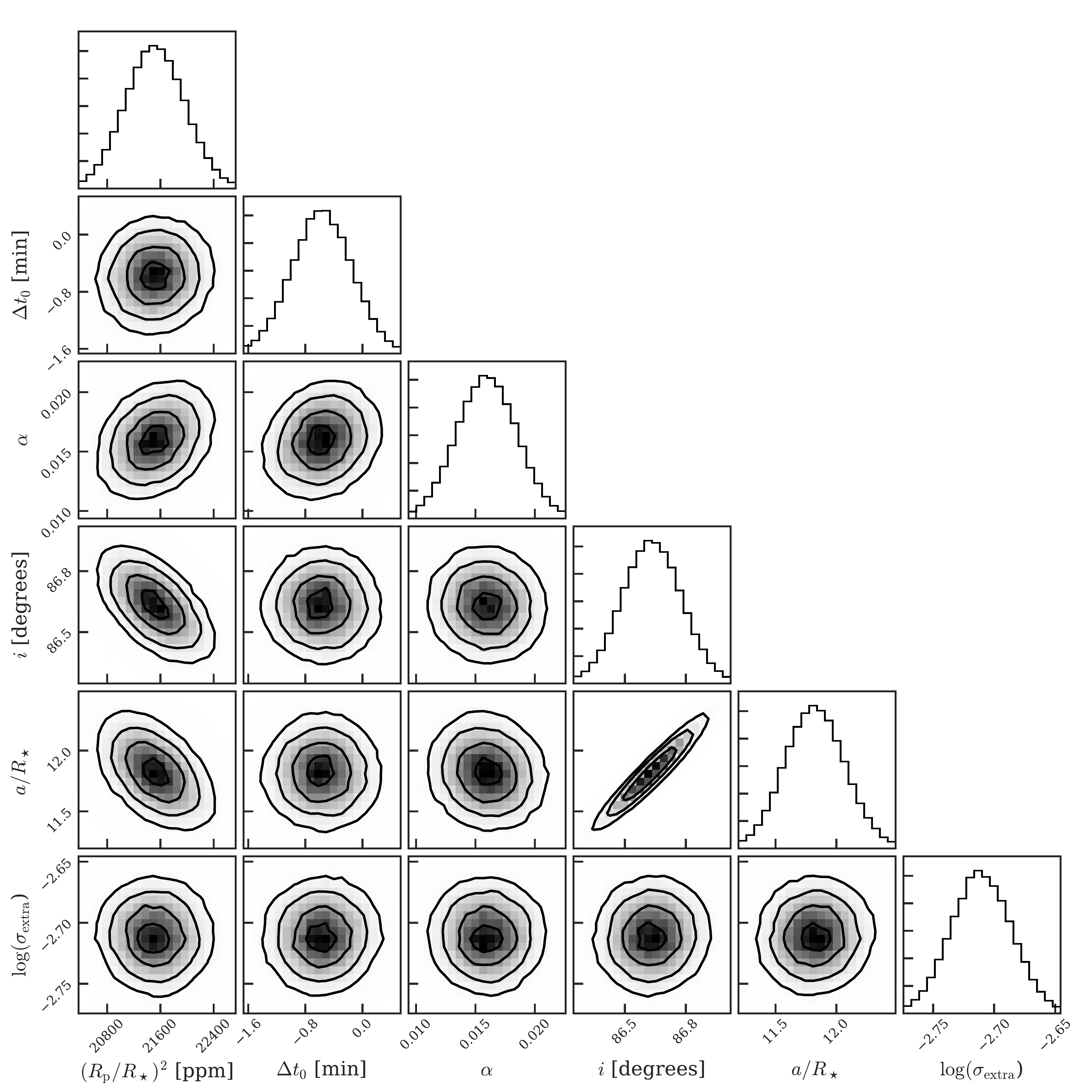}
    \caption{Corner plot of the posterior probability distributions for our fit to WASP-69b. The middle 99\% of samples are shown with contours denoting 1, 2, and 3$\sigma$ boundaries.}
    \label{fig6}
\end{figure*}

\begin{figure*}[ht!]
    \centering
    \includegraphics[width=\textwidth]{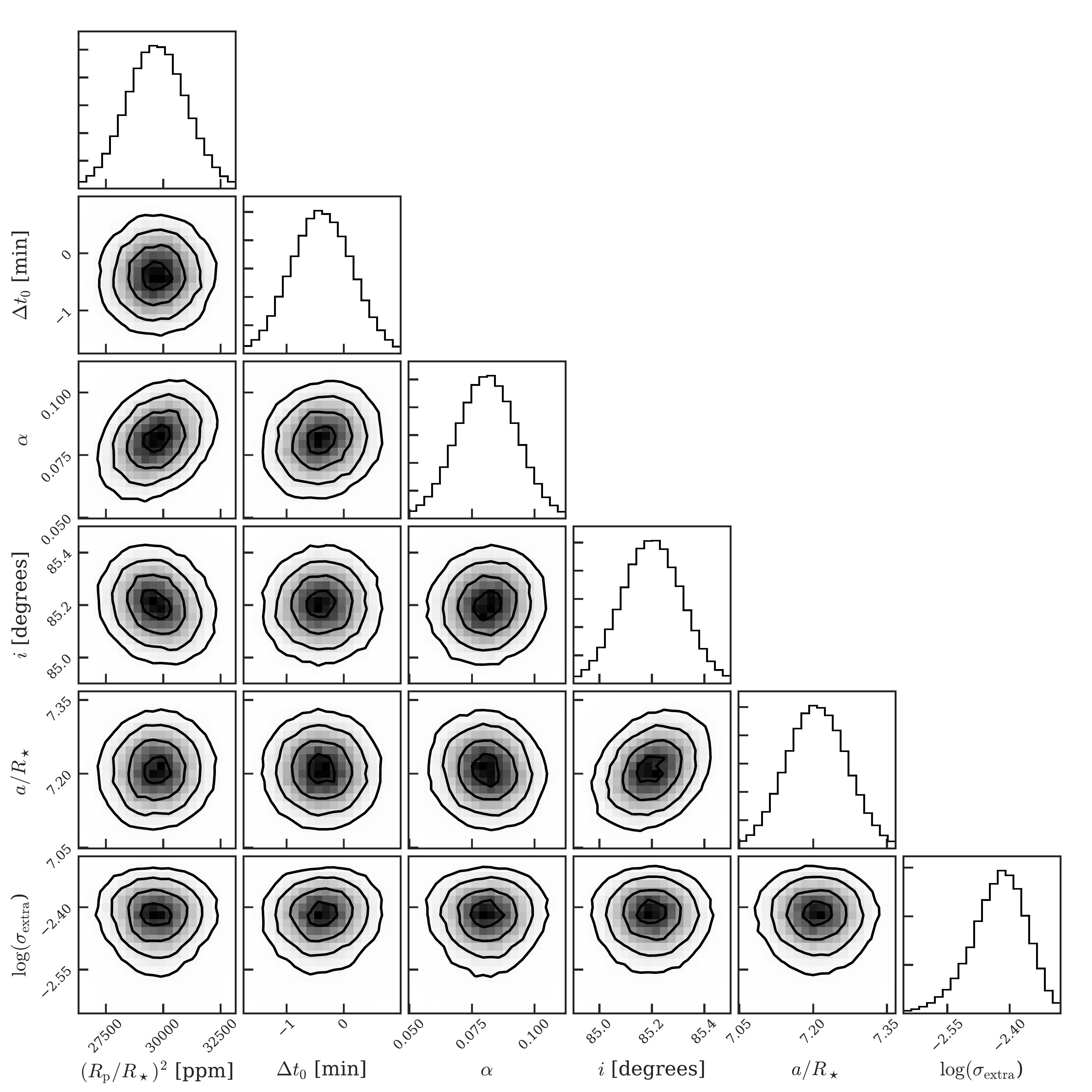}
    \caption{Same as Figure~\ref{fig6} but for WASP-52b. }
    \label{fig7}
\end{figure*}

\end{document}